%% file: llncs.tex
\documentclass{llncs}
\usepackage{makeidx}  % allows for indexgeneration
\usepackage[textsize=scriptsize]{todonotes}
\usepackage{xspace}
\usepackage{fancyhdr}
\begin{document}
\mainmatter              % start of the contributions

\cfoot{\textit{The final publication is available at link.springer.com}}

\title{Analysis of Human Awareness of Security and Privacy Threats in Smart Environments}

\titlerunning{Analysis of humans' security and privacy threat in smart environments}  % abbreviated title (for running head)
\author{Luca Caviglione\inst{1} \and Jean-Fran\c{c}ois Lalande\inst{2,3} \and \\ 
Wojciech Mazurczyk\inst{4} \and Steffen Wendzel\inst{5}}
\authorrunning{Luca Caviglione et al.} % abbreviated author list (for running head)
\institute{Institute of Intelligent Systems for Automation (ISSIA), National 
Research Council of Italy (CNR), I-16149, Genova, Italy.\\
\email{luca.caviglione@ge.issia.cnr.it}
\and
Inria, Univ. Rennes 1, Sup\'elec, CNRS, IRISA UMR 6074, F-35065 Rennes, France
\and
INSA Centre Val de Loire, Univ. Orl\'eans, LIFO EA 4022, F-18020 Bourges, France\\
\email{jean-francois.lalande@insa-cvl.fr}
\and
Institute of Telecommunications, Warsaw University of Technology, 00-665 Warsaw, Poland \\
\email{wmazurczyk@tele.pw.edu.pl}
\and
Fraunhofer Institute for Communication, Information Processing and Ergonomics (FKIE), D-53113 Bonn, Germany\\
\email{steffen.wendzel@fkie.fraunhofer.de}
}

\maketitle              % typeset the title of the contribution

\begin{abstract}
Smart environments integrate Information and Communication Technologies (ICT) into devices, vehicles, buildings and cities to offer an increased quality of life, energy efficiency and economical sustainability. In this perspective, the individual has a core role and so has networking, which enables such entities
to cooperate. 
However, the huge amount of sensitive data, social aspects and the mixed set of protocols offer many opportunities to inject hazards, exfiltrate information, mass profiling of citizens, or produce a new wave of attacks. This work reviews the major risks arising from the usage of ICT-techniques for smart environments, with emphasis on networking. Its main contribution is to explain the role of different stakeholders for causing a lack of security and to envision future threats by considering human aspects.
\keywords{privacy, security, steganography, smart buildings, human aspects.}
\end{abstract}

\section{Introduction}
Smart buildings are an elementary component of smart environments, which aim at improving the comfort of individuals and their lifestyle. In essence, they integrate \emph{Information and Communication Technologies} (ICT) into devices, vehicles and buildings to provide a higher quality of life, a reduced environmental footprint and economical benefits. Pushed to the limit, such basic blocks can be arranged to produce large-scale deployments known as smart cities. 
Smart environments are the result of a large interdisciplinary effort, ranging from civil engineering to cloud computing. However, in this work we focus on networking/devices, since they provide many important features such as: \textit{i}) the ability of collecting information from the surrounding environment, \textit{ii}) the possibility of sending remotely commands and feedback, also in a real-time fashion, \textit{iii}) the availability of an infrastructure to handle the resulting amount of data. 
Prime examples are, among the others: wireless loops used to gather information from sensors, \emph{Radio Frequency Identification} (RFID) deployed to monitor the status of a physical area, and the \emph{Internet of Things} (IoT) paradigm to access and control devices or assets like locks, light and household appliances~\cite{snoon}.

Alas, smart technologies are tightly coupled with individuals, especially in terms of their lifestyles, bad habits and sensitive data. Therefore, all the information gathered, exchanged and stored within smart environments can lead to severe issues in terms of security and privacy. For instance, detailed personal information can be used for mass profiling or social engineering attacks. Moreover, misuse of devices, bad habits and poor understanding of handled technology can lead to severe security breaches. For example, smartphones are often used to control different parts of smart environments (e.g., buildings) and worms or infected applications could attack the discovered appliances of the user's environment.

In this perspective, this paper analyzes the most relevant security and privacy threats of users in smart environments. Given the example of smart building security, we highlight the practical background which leads to a lack of security functionality and awareness at the side of vendors, integrators and operators.
The main contributions of this paper are the following: the systematic review of hazards rooted within the most relevant smart paradigms; an assessment of emerging threats arising by the mix of ICT technologies and human aspects; the discussion of possible countermeasures to mitigate identified security issues. We point out that this paper serves as an introductory work for the HAS session on \emph{Human Aspects of Information Security, Privacy and Trust for Smart Buildings}. 

The remainder of the paper is structured as follows: Section \ref{ps} describes the problem space generated by the smart paradigm. Section \ref{sb} deals with human aspects of insecurity of smart buildings, Section \ref{sph} concentrates on smartphones, and Section \ref{sv} reviews vehicles. Section \ref{haw} proposes a role-based perspective on threats related to smart environments and Section \ref{fv} gives our future vision of identified threats and possible countermeasures.  Section \ref{concl} concludes the paper. 

\input{pb.tex}

\input{network_threats.tex}

\input{smart_threats.tex}

\input{future.tex}

\section{Conclusion}
\label{concl}

This paper discussed the human-related security and privacy aspects of smart environments. We highlighted the resulting consequences for humans when various attacks on smart buildings, smart phones, and smart vehicles are performed, also by emphasizing the role of inter-connected things populating smart cities. 
%-- things, which are inter-connected in smart cities.
%TBA: more text here. any ideas?
Furthermore, by discussing the example of smart buildings, we conclude that awareness for smart things is a multifaceted problem. Vendors, customers, and operators as well as awareness for the deployment process of smart things must be considered. We pointed out that research efforts have already been started from smartphones but remain very limited for vehicles, devices and especially for buildings.

We also conclude that a variety of attacks will be possible in a near future. Therefore, we underline the importance of a rapid development of proper and effective countermeasures.
Possible countermeasures have to be inspired by the efforts achieved in other fields like peer-to-peer, ad-hoc networks and regular computers. The customer's comprehensiveness of these future security measures is a central requirement in order to be effective. This is a prime research task both for the academia and the industry in order to improve the security of smart environments.

\bibliographystyle{splncs03}
\bibliography{llncs}

\end{document}

%% file: pb.tex
\section{Description of the Problem Space}
\label{ps}

Smart environment is an umbrella term comprising different kinds of devices or specific deployments. As today, the most relevant areas of smart things are:

\begin{itemize}
\item \textbf{Smart Buildings:} they collate a mix of smart devices as to produce an integrated environment. For such a complex deployment, proper middleware in charge of offering a coherent access is usually adopted, as well as proper computing facilities to store user data, process control directives and provide optimizations, especially in the field of energy management. Thus, this scenario can be used to exploit data hiding, for instance to covertly orchestrate a botnet.

\item \textbf{Smart Devices:} they are quickly becoming widespread and one of the core blocks to pursue the vision of a more ``human-centric" environment. Common examples of smart devices are gaming consoles, set-top-boxes, light bulbs and household appliances. These devices can be used to infer habits, even political views, for instance by evaluating the shows watched on the TV. Moreover, the availability of full-featured TCP/IP stacks can be exploited to produce new types of botnets in order to e.g. amplify spam campaigns. It must be emphasized that smart devices will not be investigated in details in this paper due to space limitations and their technical heterogeneity.

\item \textbf{Smart Phones:} are the most popular tools used to interact with other devices, and can be used to remotely control buildings and vehicles. In addition, they are the preferred platform to connect to the Internet and to communicate using heterogeneous networks (e.g., cellular or WiFi). They store a huge source of sensitive details such as messages and contacts and can be paired with smart watches which increases their potential to collect personal data. As a result, they are one of the preferred targets for data exfiltration of users, while additionally empowering phishing, social phishing, cyber bullying and social engineering~\cite{cavcoc}.

\item \textbf{Smart Vehicles:} modern automotives offer features to geolocate vehicles, mainly through the \emph{Global Positioning System} (GPS), and to plan routes. Such features are not only used by individuals, since they are at the basis of fleet management and intelligent/smart transportation services. Also, modern vehicles can remotely send telemetry data as to prevent fault and guarantee proper service levels, e.g., for  goods delivery. This allows massive user profiling, which leads to understand habits to conduct physical attacks.

\end{itemize}

The resulting problem space is very composite and needs a thorough understanding of all the technical components used, both to evaluate the degree of (in)security and to engineer proper countermeasures and mitigation techniques. To this aim, functional entities needing an investigation are: \textit{i}) wireless networks (e.g., the IEEE 802.11) as well as the core protocols used to exchange data or grant human interaction (e.g., HTTP); \textit{ii}) elaborate a proper taxonomy/ranking to understand where the related weaknesses impact more (e.g., physical security vs.\ cybersecurity); \textit{iii}) understand how sensitive data can be used also jointly with those available on \emph{Online Social Networks} (OSNs) as a method to produce a new wave of attacks; \textit{iv)} understand why standard detection methods can be defeated by the complexity and diversity of smart  environments.

%% file: network_threats.tex
\section{Smart Buildings}
\label{sb}

Smart buildings are automated buildings, i.e., those comprising \emph{Building Automation Systems} (BAS) and inter-connected with the IoT. The importance of BAS for today's societies increases steadily due to various reasons. For instance, being enriched with more features, buildings can perform an additional number of routine tasks such as energy saving and in the context of an aging society, smart buildings ensure that elders can stay longer in their homes before being forced to move to a nursing home.

Various vulnerabilities in the available standards of communication protocols used in BAS are known. Most of these communication protocols, e.g., EIB/KNX, LON or BACnet, were designed many years ago with very limited focus on IT security~\cite{SecNetBAS06,SecBAS10}. Improved standards for the most-widely used BAS protocols are already proposed or under development, however, the application of these enhanced protocols in practice and the integration into products is currently not present. Moreover, the integration of newer protocols into legacy BAS environments is hardly feasible and thus, novel solutions like \textit{traffic normalization} must be applied which protect legacy systems~\cite{Szl:BACnetTN}. From a human-oriented perspective, the major attacks which can be performed on buildings are:

\begin{itemize}
 \item \textbf{Surveillance}: as shown in~\cite{Mundt:Wash,Wendzel:3SL}, it is technically feasible to perform surveillance of events in buildings, e.g., caused by inhabitants or employees. Therefore, passive and active attacks are known: the attacker either exploits side channels or directly requests sensor values from Internet-connected BAS. An attacker can, for instance, use surveillance to monitor the behavior of inhabitants or employees.
 
 \item \textbf{Remote control}: while surveillance relies on sensor values and actuator states in a smart building, a remote control is feasible, too. Therefore, actuators are used to perform actions, triggered by the attacker. For instance, to break into a building, a thief can send commands to window actuators/door actuators and can attack the physical access control system. In worst case, remote control attacks influence the safety of inhabitants and people working in a building.
 
 \item \textbf{Physical exploitation}: being a form of remote control, an attacker can at least indirectly get advantage of exploiting the BAS of other households. For illustration, we use the example of a house with two parties A and B, each possessing its own flat and own BAS. Imagine party A leaves his flat, which is underneath the flat of B, for winter holidays while party B is staying at home. Since the ceiling of A's flat is the ground floor of B's flat, B can attack the BAS of A to maximize the heating level in A's flat. As a result, the temperature of B's flat is also heated a little what saves heating costs for B while increasing the costs for A.
 
 \item \textbf{Availability}: the functioning of a building is essential for today's organizations. Hence, causing a \emph{Denial of Service} (DoS) attack, e.g., by simple misconfiguration, can affect all areas of building automation, such as physical access control or fire alarm systems, and is thus not only harmful for enterprise processes but for the safety of people.
 
 \item \textbf{Smart Building Botnets}: when surveillance or remote control is not only performed for a single household or industrial building, but for a larger number of buildings, novel scenarios emerge. So-called \emph{smart building botnets} can perform mass surveillance and mass remote control~\cite{Wendzel:SBB}. For instance, a local oil distributor may rise his sales by slightly increasing the nightly heating levels of his customer's households. Such large-scale attacks potentially influence the privacy, safety, and living of inhabitants and employees in whole regions.
\end{itemize}

\section{Smart Phones}
\label{sph}
For years, smartphones have been one of the most important tools to communicate and store personal data. Recently, the advent of frameworks for managing home appliances, monitoring the health of the owner and handling payments, led to an important paradigm shift. In essence, smartphones are the preferred dashboard to access smart homes, and interact with appliances or vehicles. In addition, the security is under the responsibility of the user, since he/she is in charge of managing the administration and installation of the applications, as well as of undertaking security decisions. Yet, there is not any guarantee about his/her level of technical knowledge, which makes the human an effective vector of attack. As a consequence, smartphones are prone to different attacks in terms of human aspects, specifically:

\begin{itemize}
\item \textbf{Data exfiltration}: capturing user's personal data is one of the primary goals of attackers~\cite{Felt2011b}. After collecting the data, for instance via phishing, the malware uploads it to a remote server. Thus, personal or business information is not only stolen but also stored in a place inaccessible for the user.
\item \textbf{Exploitation of acquired resources}: a classical secondary goal consists of exploiting the controlled smartphone~\cite{Felt2011b}. For example, it can be used as: a client of a botnet network, to send premium-rate SMS, or to participate in computations for mining bitcoins. From the user perspective, it disturbs the normal behavior of the smartphone and can result in additional cost. As other smart entities can be controlled by an application on the user's smartphone, compromising the smartphone can also give a fresh starting point to attack other smart entities on the same network.
\item \textbf{Surveillance}: a malware can try to access user's localization and report these data to the attacker. Using the collected positions, more complex attacks can succeed to infer the identity of the user~\cite{gambs}.
% Added into 'exploit acquired resources' section as it is not so much related to "human" and not so much recognised in the literrature.
%\item \textbf{Pivoting or island hopping}: as other smart entities can be controlled by an application on the user's smartphone, compromising the smartphone gives a fresh starting point to attack other smart entities on the same network. If needed, the attacker's malware can ask for extra credentials to the user through the smartphone when a particular smart device is detected on the local network.
\item \textbf{Battery drain}: a severe fragility of smartphones is the intrinsic power limitation due to the usage of battery. Hence, its malicious depletion can be at the basis of a new kind of DoS %DoS already introduced, thus not writing full term here --steffen
attack, where the device is made unusable~\cite{bdrain}. Possible mechanisms range from the injection of energy wasting code within a malware or a stimulation of the device via its air interfaces. In any case, this attack may isolate the victim, making communications with the rest of the world infeasible. This kind of hazard is also effective for the case of sensors or nodes of an IoT deployment and the victim could have his/her safety framework compromised.

\item \textbf{Information hiding}: as a consequence of a full implementation of the TCP/IP protocol stack, and the diffusion of BAS over IP solutions, modern smartphones run several applications using a very mixed set of protocols. The latter can be exploited for information hiding purposes. In essence, multimedia data and network traffic can be used as legitimately appearing carriers by information hiding techniques to make a third-party observer unaware of the resulting flow. This technique can be also used for mass-profiling~\cite{wenpap} or for empowering malware exfiltrating data~\cite{wojsur}.
\end{itemize}

\section{Smart Vehicles}
\label{sv}
The most useful scenario envisaged for smart vehicles concerns \emph{Vehicular Ad-hoc Networks} (VANETs), which offer features such as, road safety, route planning, entertainment, tolling, traffic management and support for intelligent transportation. 
In such a deployment, humans should be not endangered by vehicles as a vehicle's misbehavior can lead to severe injuries and safety hazards\footnote{Def Con 21 talk by C. Miller and C. Valasekentitled entitled ``Adventures in Automotive Networks and Control Units" \textit{https://www.defcon.org/html/defcon-21/dc-21-speakers.html\#Miller}}$^,$\footnote{Def Con 18 talk by M. Metzger entitled ``Letting the Air Out of Tire Pressure Monitoring Systems" \textit{https://www.defcon.org/images/defcon-18/dc-18-presentations/Metzger/DEFCON-18-Metzger-Letting-Air-Out.pdf}}. In addition, the cooperative nature of VANETs puts the network in a central and critical role. Therefore, networking technologies used in vehicles must be protected against malicious activities, which are very effective~\cite{Koscher2010,Checkoway2011}. Their main scopes are to propagate incorrect information about events on the road, to gain sensitive information, and to disrupt the network infrastructure to prevent users accessing the service~\cite{icnit}.
Among the others, the most relevant attacks in terms of human aspects are: 

\begin{itemize}
\item \textbf{Injecting bogus information}: an attacker deliberately injects false information into the network to produce arbitrary situations along a route \cite{Al-kahtani2012}. As an example, a node sending false information to benefit from a reduction of traffic along a common path. This can be done via false information reporting traffic jams, road accidents, and blocked routes as to suggest alternative ways. The most popular methods to achieve such goals are: intentionally creating or modifying existing frames, repeating previously captured data (replay attack), and misleading vehicle's sensors (illusion attack).
% For example, the latter can inform that the car has been involved in an accident or is under bad weather conditions. 
%From the user perspective it leads to misinformation that results in e.g. longer traveling time due to e.g. taking detour despite the non-existent accident broadcasted by an attacker.

\item \textbf{Sybil attack}: it is based on spoofing the identity of nodes to flood the network with incorrect information \cite{Zeadally2012}. Typically, the attacker produces multiple copies of false data to appear as legitimate. Then, such false data can be used to induce the same reactions previously explained. 
%
%This can become a serious threat when a single node is able to force other legitimate vehicles to accept forged messages and to treat them as legitimate ones. The typical Sybil attacks produce multiple copies of false data to appear as legitimate. For non-malicious users the results due to successful Sybil attack are similar like in the previous case.

\item \textbf{Wormhole attacks}: it creates a tunnel between two attackers' vehicles as a way to inject false information and disrupt the vehicular network \cite{Al-kahtani2012}. 
The wormhole attack can be especially dangerous since it makes routing tables
incoherent, thus causing the unavailability of the service for humans. 
%the lack of access to the smart 
%
%for the typical routing algorithms such as the AODV (Ad-hoc On Demand) and the DSR (Dynamic Source Routing). This attacks can cause serious vehicle network disorganization and result in significant harm for humans due to e.g. VANET infrastructure improper functioning. 

\item \textbf{Routing protocols attacks}: vehicular networks use a mixed amount of broadcast and multi-hop traffic, e.g., to deliver data to isolated nodes. In this case, the injection of bogus routing information can cause the routing protocols to misbehave~\cite{blackhole,vanet_route}. This may lead to the extension of packets' routes, the creation of routing loops, or the redirection of the traffic to an unreal node (blackhole attacks) or towards the attacker (greyhole attacks). 
%The result from a legitimate user's perspective is similar as in the previous case.
%
%
%
% most of the traffic delivered through the vehicular network is broadcast, yet a fraction could rely multi-hop methods, for instance to serve partially isolated entities. Besides typical passive attacks that focus on gathering valuable information, e.g., network topology, location of the nodes and/or their role in the network, more sophisticated active attacks are possible. For instance, by injects bogus routing information into the network thus causing the routing protocols to misbehave. This may lead to: extension of packet route, creation of routing loops, redirection of the traffic to an unreal node (blackhole attacks) or towards the attacher (greyhole attacks)~\cite{blackhole},~\cite{vanet_route}. The result from legitimate user perspective is similar like in the previous case.

\item \textbf{Man in the middle attacks}: there are no significant differences between Man in the middle attacks in VANETs and in typical wired networks. These attacks impact the authenticity of transmitted information which may threaten the privacy and identity of users.

\item \textbf{DDoS attacks}: two main techniques are utilized to perform DoS attacks \cite{Zeadally2012}. First, by disrupting the frequencies utilized for wireless communication, the attacker produces a jam in a given frequency range. This is quite easy to implement and its effectiveness mainly depends on the transmitting power of the jamming device. Second,  by sending large amounts of network traffic by an authorized host, the attacker generates network messages that are valid but with high volumes/rates thus causing congestion, latency and intermittent connectivity. In both cases, the vehicle is unable to send/receive any information and the driver could potentially miss an important announcement, e.g. on the accident nearby or the worsening of weather conditions.

\item \textbf{GPS spoofing}: it is based on sending a ``louder" GPS signal to hide the legitimate one \cite{Zeadally2012}. Current protection methods are mainly based on monitoring the power expected by a legitimate GPS satellite. This attack may lead to a car accident or to driving a vehicle in an abandoned area.

\end{itemize}

%% file: smart_threats.tex
\section{Human Awareness of Smart Environment's Threats: a Role-based Perspective}
\label{haw}

When comparing human aspects of ICT-related topics with those in smart environments, a clear difference can be recognized. In ICT-related areas, a strong development of security features is achieved. Thus, vendors integrate security into their products and customers clearly demand for such features. In smart environments, especially the classic ones -- such as factory automation -- there is a clear lack of security, which we mainly illustrate in this section by reviewing the point of view of the different actors in the case of BAS. Most of these views were obtained owing to personal conversation with the different stakeholders.

\paragraph{Vendors.} Vendors do not integrate security into their automation equipment as they lack know-how. They focus on engineering aspects and product quality is rather measured in longevity of components instead of in terms of security. When security features are integrated, these are in many cases implemented from scratch. For instance, a number of German BAS vendors promote their BAS network components explicitly with the ``feature'' that instead of buying a network stack from a country abroad, they have one competent engineer who implemented the stack himself. A one-person implementation of a network stack, such as BACnet, including its complex features is hardly feasible in a secure way by a single engineer.

\paragraph{Customers.} On the other hand, the customers lack security awareness as well. Awareness-raising processes are currently taking place on a regional, national, and international scale. For instance, the 28th German \textit{GLT Anwendertagung} -- a leading event for professional customers --
%\footnote{\textit{http://www.glt-anwendertagung.de/} (in German)}, a national event, ===> we cannot cite a work not in english...
organized a security session on BAS in 2014. %, and the NIS Platform\footnote{\textit{https://resilience.enisa.europa.eu/nis-platform}} announced in the Cybersecurity Strategy of the European Union features a working group which creates a \textit{Strategic Research Agenda}, which covers security aspects for smart buildings as one of their topics.
However, the effect of these awareness raising processes is small. For this reason, possessing still no (or very limited) awareness for security threats, customers do not demand security features from \textit{vendors}, which, in turn, see the implementation of additional security features as costly.

\paragraph{Operators.} Operators of smart buildings are usually janitors without any know-how on the IT security of their BAS. Even if the operators received additional education on BAS (e.g., certificates on building management), these courses lack any security features. The perspective of an operator is to ensure the functioning of a BAS, including its \textit{safe} operation, for instance an intact fire alarm system, but security aspects are considered an additional overhead.

Additionally, \textit{vendors} provide no tools to monitor or configure the security of BAS components and thus, even if \textit{operators} would possess knowledge on IT security, they could not apply it in practice.
In particular, as smart environments are in most cases networked environments, operators require \textit{cyber situational awareness} \cite{SitAwarenessSurvey14}, for example the awareness of any kind of suspicious activity taking place in cyberspace. In various cases, such as larger or inter-connected BAS, the number of events cannot be processed by human operators without any support. To this end, research came up with visualization approaches, which, for instance, present information in such a way that events with higher entropy are easier to spot. %~\cite{Wendzel:FutureSec14}.
However, in practice, these tools are not available and if available are used for spotting misconfiguration problems or malfunctioning equipment instead of detecting cyber security-related attacks. 
%In other words, even if a cyber security event would occur in an advanced visualization system, the chances are that operators misunderstand these events and will search for configuration or malfuntioning components instead of countering an attack.

\paragraph{Project deployment.} Construction of a BAS suffers from non-optimized information exchange of the parties involved in the design, construction, and operation process of a building \cite{Noeldgen:FutureSec14}, which includes the planning, integration, and operation of a BAS. Moreover, know-how about the operation must be managed, including to consider its potential loss if operators leave the organization---a problem that is even more important for other critical smart areas such as operator centers for naval vessels or railways \cite{Bronkhorst:FutureSec14}.

\medskip

This analysis is particularly pessimistic for BAS. Other smart components considered in this paper have made better security efforts.
We summarize human awareness from a security perspective for each ``component" of the smart environment in Table~\ref{sum_smart}. Smartphones have received better attention than vehicles or devices. They benefit from two effects: customers ask for more security because of the increasing connectivity with OSNs; vendors can integrate adapted security technologies that have been matured for GNU/Linux operating systems, especially since Android has taken the lead in the market.

\begin{table*}[tb] 
\caption{Summary of human awareness from a role-based perspective for each "component" of smart environment.}
\begin{center}
\begin{tabular}{| p{3.5cm}| p{2cm}| p{2cm}| p{2cm}| p{2cm}|}
\hline
\textbf{Smart ``component"} & \textbf{Vendors} & \textbf{Customers} & \textbf{Operators} & \textbf{Deployment}\\
\hline
\hline
Buildings & Low & Low & Low & Low  \\
\hline
Vehicles & Medium & Low & Low & Low  \\
\hline
Phones & High & Medium & High & Medium  \\
\hline
Devices & Medium & Low & Low & Low  \\
\hline
\end{tabular}
\end{center}
\label{sum_smart}
\end{table*}

%% file: future.tex
\section{Future Vision on Threats and Countermeasures}
\label{fv}

Today, a number of the attacks presented in previous sections, e.g., smart building botnets, should rather be considered technically feasible than a real-world threat. However, given the linked risks for individuals and communities and the lack of awareness of the involved roles, the hurdles for attackers are considered not higher than for other ICT attacks. For this reason and since smart things of each type quickly gain more widespread, authors who discussed the particular attacks conclude the importance of a rapid countermeasure development as potential attacks are known before emerging on a larger scale in practice (e.g. \cite{Wendzel:SBB}).

\subsection{Future Threats}
The potential of attacks can be considered larger if already known attacks from other areas of IT security are getting adapted to smart things. For instance, \emph{watering hole attacks}~\cite{Lowe2014} can be adapted to smart buildings/smart phones. Consider a community that is living in the same building. If an attacker wishes to access the BAS it is enough for her to infect only one inhabitant's smartphone which she uses to control the smart building and eventually other habitants will be infected. This scenario becomes even more significant as some hotels announced to enable smartphone-based hotel room access for guests.

In this perspective, smartphones will definitely be one of the preferred playgrounds to exploit threats. Especially, this is due to the complexity of their security policies, which discourages users to analyze and take adequate decisions. In addition, smartphones possess authentication tools that become of high interest for attackers. %striked because does not say much: Therefore, this contradiction can lead to severe information leakage or the gain control of applications or devices. 

One of the examples of how future mobile malware can covertly exfiltrate user's sensitive data is envisioned in~\cite{istegsiri}. The proposed steganographic method takes advantage of the built-in Siri service which has been offered for iPhone/iPad as a native service from iOS5 in 2011. Siri allows interacting with the iOS-based device using voice commands. To offload the device, the translation of voice inputs to text is performed remotely in a server farm operated by Apple. To this aim, the iPhone/iPad samples the voice, sends it to a remote facility, and waits for a response containing the recognized text, a similarity score and a time stamp. This characteristic feature can be exploited by an attacker which could produce ad-hoc voice patterns to manipulate the throughput and encode a secret into its shape. In future, a similar approach can be applied to all services relying on a massive conversation between the user's device and  similar services in the cloud like GoogleVoice for Android OS or Cortana for Windows Phone.

For other smart devices, the potential of attack is dramatically increasing. For example, the Rapid7 company published in 2013 a security report about several critical vulnerabilities of the UPnP library~\cite{rapid7-2013}. These vulnerabilities affect billions of devices, for example Smart TVs, and gives opportunities to build attacks and gain root shells on these devices.

Lastly, because smart devices typically reside inside smart buildings, a compromised device  will help an attacker to attack smart buildings.
% Due to space limitation and their fast evolving nature, we omit here a thorough analysis, which is partially overlapped with more ``classical" network appliances. In fact, an attacker wanting to force a smart devices usually search for vulnerabilities in the firmware or in the implementation of the protocol stack. 
The attacker can try to capture data, infer residents habits, such as food products ordered online (smart fridge) and TV shows watched (smart TV). This can significantly impact privacy and enable the production of a new wave of extremely precise (and effective) social engineering attacks.

\subsection{Future Countermeasures}

A number of futuristic protection approaches for smart things are imaginable. 
%
%For smart vehicles the following scenario that would surely improve security on roads can be envisioned. Consider a vehicle capable to verify whether the driver is drunk/tired. If such a case occurs then a vehicle can react by first sending some warning messages to the driver and in a second stage prevent him from driving. And even if this fails, then a smart vehicle can warn other road participants about the potential threat.
%
For smart vehicles, the used protocols should include validation algorithms as it is clear that there, many potential opportunities arise for an attacker to inject malicious information. 
As used protocols should react in a real-time fashion, the added security should be lightweight and distributed between participants in order to give robust results. These solutions, reviewed in~\cite{Engoulou20141}, can be based on reliable cryptographic key distribution and has been already actively studied for example for ad-hoc networks. With such tools, the privacy of users should be guaranteed. Also, they can be based on the reputation systems already deployed for peer-to-peer architectures. 

A mean for smart buildings could be to introduce multilevel security \cite{Wendzel:3SL}. Such an approach could, for instance, prevent that devices in a storage room could read sensor values from the management floor of an organizational building.

%JFL: ??
%Another important factor that will significantly impact the effectiveness of countermeasures is to spread security and privacy awareness among different roles linked to smart environments (vendors, customers, operator, etc.).

For the attack vectors discussed for smartphones, the industry is currently working on \emph{Trusted Execution Environments} (TEE) that would introduce a secured trusted space of execution while the regular operating system remains untrusted. This way, vendors would be able to split their applications and protect the critical parts into the smartphone's TEE~\cite{Arfaoui2014}. Moreover, malware detection is one of the hot topics for researchers in mobile security. Nevertheless, current anti-malware products are easily defeated by transformation techniques of the malware's code~\cite{Rastogi2013}. Thus, these aspects remain to be addressed.

%% file: llncs.bbl
\begin{thebibliography}{10}
\providecommand{\url}[1]{\texttt{#1}}
\providecommand{\urlprefix}{URL }

\bibitem{Al-kahtani2012}
Al-kahtani, M.: Survey on security attacks in vehicular ad hoc networks
  (vanets). In: Signal Processing and Communication Systems (ICSPCS), 2012 6th
  International Conference on. pp. 1--9 (Dec 2012)

\bibitem{Arfaoui2014}
Arfaoui, G., Gharout, S., Traor\'{e}, J.: {Trusted Execution Environments: A
  look under the hood}. In: The International Workshop on Trusted Platforms for
  Mobile and Cloud Computing. pp. 259--266. IEEE Computer Society, Oxford, UK
  (Apr 2014)

\bibitem{Bronkhorst:FutureSec14}
Bronkhorst, A., Post, W., te~Brake, G.: From human factors to {HSI} and beyond:
  Design of operations centers and control rooms. In: 9th Future Security --
  Security Research Conference. pp. 140--146. MEV Verlag (September 2014)

\bibitem{cavcoc}
Caviglione, L., Coccoli, M.: Privacy problems with web 2.0. Computer Fraud \&
  Security  2011(10),  16 -- 19 (2011)

\bibitem{istegsiri}
{Caviglione L., Mazurczyk W.}: Understanding information hiding in {iOS}. IEEE
  Computer magazine  (January/February 2015)

\bibitem{Checkoway2011}
Checkoway, S., McCoy, D., Kantor, B., et~al.: Comprehensive experimental
  analyses of automotive attack surfaces. In: Proceedings of the 20th USENIX
  Conference on Security. pp. 6--6. SEC'11, USENIX Assoc., Berkeley, CA, USA
  (2011)

\bibitem{Engoulou20141}
Engoulou, R.G., Bellaïche, M., Pierre, S., Quintero, A.: {VANET} security
  surveys. Computer Communications  44(0),  1 -- 13 (2014)

\bibitem{Felt2011b}
Felt, A.P., Finifter, M., Chin, E., Hanna, S., Wagner, D.: {A survey of mobile
  malware in the wild}. In: 1st ACM workshop on Security and privacy in
  smartphones and mobile devices. p.~3. ACM Press, New York, New York, USA (Oct
  2011)

\bibitem{SitAwarenessSurvey14}
Franke, U., Brynielsson, J.: Cyber situational awareness -- a systematic review
  of the literature. Computers \& Security  46,  18--31 (July 2014)

\bibitem{gambs}
Gambs, S., Killijian, M.O., Nunez~del Prado~Cortez, M.: De-anonymization attack
  on geolocated data. In: Trust, Security and Privacy in Computing and
  Communications (TrustCom), 2013 12th IEEE Int. Conf. on. pp. 789--797 (2013)

\bibitem{SecNetBAS06}
Granzer, W., Kastner, W., Neugschwandtner, G., Praus, F.: Security in networked
  building automation systems. In: Factory Communication Systems, 2006 IEEE
  International Workshop on. pp. 283--292 (2006)

\bibitem{SecBAS10}
Granzer, W., Praus, F., Kastner, W.: Security in building automation systems.
  Industrial Electronics, IEEE Transactions on  57(11),  3622--3630 (November
  2010)

\bibitem{Koscher2010}
Koscher, K., Czeskis, A., Roesner, F., Patel, S., Kohno, T., Checkoway, S.,
  McCoy, D., Kantor, B., Anderson, D., Shacham, H., Savage, S.: Experimental
  security analysis of a modern automobile. In: Security and Privacy (S\&P),
  2010 IEEE Symposium on. pp. 447--462 (May 2010)

\bibitem{icnit}
Lipi{\'n}ski, B., Mazurczyk, W., Szczypiorski, K., {\'S}mietanka, P.: Towards
  effective security framework for vehicular ad-hoc networks. In: Proc. of 5th
  International Conference on Networking and Information Technology (ICNIT
  2014) (2014)

\bibitem{Lowe2014}
Lowe, M.: {Defending against cyber-criminals targeting business websites}.
  Network Security  2014(8),  11--13 (Aug 2014)

\bibitem{vanet_route}
Lu~Chen, Hongbo~Tang, J.W.: Analysis of {VANET} security based on routing
  protocol information. In: Proc. 4th Int. Conf. Intelligent Control and Inf.
  Proc. (2013)

\bibitem{bdrain}
Martin, T., Hsiao, M., Ha, D.S., Krishnaswami, J.: Denial-of-service attacks on
  battery-powered mobile computers. In: Pervasive Computing and Communications,
  2004. PerCom 2004. Proceedings of the Second IEEE Annual Conference on. pp.
  309--318. IEEE (2004)

\bibitem{wojsur}
Mazurczyk, W., Caviglione, L.: Steganography in modern smartphones and
  mitigation techniques. Communications Surveys Tutorials, IEEE  PP(99),  1--1
  (2014)

\bibitem{rapid7-2013}
Moore, H.: {Security Flaws in Universal Plug and Play}. Tech. Rep. January,
  Rapid7 (2013), \url{https://community.rapid7.com/docs/DOC-2150}

\bibitem{Mundt:Wash}
Mundt, T., Kruger, F., Wollenberg, T.: Who refuses to wash hands? privacy
  issues in modern house installation networks. In: Proc. 7th Int. Conf.
  Broadband, Wireless Computing, Communication and Applications. pp. 271--277
  (November 2012)

\bibitem{Noeldgen:FutureSec14}
N{\"o}ldgen, M., Bach, A., Heinz, T.: Integration of resilience engineering in
  the trans-disciplinary building design process. In: Proc. 9th Future Security
  -- Security Research Conference. pp. 125--132. MEV Verlag (September 2014)

\bibitem{Rastogi2013}
Rastogi, V., Chen, Y., Jiang, X.: {Evaluating Android Anti-malware against
  Transformation Attacks}. In: 8th ACM SIGSAC symposium on Information,
  computer and communications security. pp. 329--334. ACM Press, Hangzhou,
  China (2013)

\bibitem{snoon}
Snoonian, D.: Smart buildings. Spectrum, IEEE  40(8),  18--23 (Aug 2003)

\bibitem{blackhole}
Subir~Biswas, Jelena~Misic, V.M.: Performance analysis of black hole attack in
  vanet. In: Proc. of 31st Int. Conf. Distributed Computing Systems (2011)

\bibitem{Szl:BACnetTN}
Szl{\'o}sarczyk, S., Wendzel, S., Meier, M., Schubet, F., Kaur, J.: Towards
  suppressing attacks on and improving resilience of building automation
  systems -- an approach exemplified using {BACnet}. In: Proc. Sicherheit 2014.
  pp. 407--418. GI (2014)

\bibitem{Wendzel:3SL}
Wendzel, S., Kahler, B., Rist, T.: Covert channels and their prevention in
  building automation protocols - a prototype exemplified using {BACnet}. In:
  Proc. 2nd Workshop on Security of Systems and Software Resiliency. pp.
  731--736. IEEE (2012)

\bibitem{Wendzel:SBB}
Wendzel, S., Zwanger, V., Meier, M., Szl{\'o}sarczyk, S.: Envisioning smart
  building botnets. In: Proc. Sicherheit 2014. LNI, vol. 228, pp. 319--329. GI
  (March 2014)

\bibitem{wenpap}
Wendzel, S., Mazurczyk, W., Caviglione, L., Meier, M.: Hidden and
  uncontrolled--on the emergence of network steganographic threats. In: ISSE
  2014 Securing Electronic Business Processes, pp. 123--133. Springer (2014)

\bibitem{Zeadally2012}
Zeadally, S., Hunt, R., Chen, Y.S., Irwin, A., Hassan, A.: Vehicular ad hoc
  networks ({VANETS}): status, results, and challenges. Telecommunication
  Systems  50(4),  217--241 (2012)

\end{thebibliography}
